# Wi-Fi & WiMAX: A Comparative Study

**Sourangsu Banerji[1*], Rahul Singha Chowdhury[2]**
[1, 2] Department of Electronics & Communication Engineering,
RCC-Institute of Information Technology, India

**ABSTRACT: Usually broadband wireless access networks are considered to be enterprise level networks providing us with more capacity as well as coverage. We have seen that in remote inaccessible areas wired networks are not at all cost effective. Wireless networking has offered us an alternative solution for such problem of information access. They have definitely changed the way people communicate and share information among themselves by overcoming problems nowadays associated with distance and location. This paper provides a comparison and technical analysis of alternatives for implementing last-mile wireless broadband services. It provides detailed technical differences between 802.11 (Wi-FI) wireless networks with 802.16 (WiMAX), a new technology that solves many of the difficulties in last-mile implementations.**

## 1. Introduction

Recently broadband wireless access (BWA) is increasingly gaining in popularity as an alternative in last mile implementations to DSL lines and cable modems. Following the successful deployment of the 802.11 wireless local area network standard, the deployment of the 802.16d and 802.16e wireless metropolitan area networks are currently in progress. The following technologies aim to provide broadband wireless access to residential areas and small business enterprises, in addition to providing internet access in countries without any wired network infrastructure. Now wireless technologies needs to be examined from two perspectives: access network and backhaul network. Even the access networks are principally categorized into two groups. There is the cellular network group and the broadband wireless access group, popularly known as the 802.xx family. The 802.xx family has seen the proliferation of different standards since its inception (Chan, 2005).

Such type of diversity is necessary due to a number of properties desired from them. The desired properties include range, bandwidth, costs of deployment, and time taken to complete deployment. Range determines the maximum area that can have full coverage. As more and more network applications emerge, bandwidth becomes critical to network efficiency. Different network standards have been developed to provide needed bandwidth. Bandwidth of a network is responsible for a number of Quality of Service (QoS) attributes that the network exhibits. Deploying a network is associated with a lot of costs. These costs determine the viability of a project and efforts are directed to balance the investment tradeoffs. Cost of deploying infrastructure is proportional to the time taken to complete deployment. For that reason, different IEEE 802.11x (Wi-Fi) and IEEE 802.16x (WiMAX) standards have been developed.

This paper is organized to describe the standards and technology associated with two of the various wireless technologies usage models. The paper then gives a technical overview of the two popular broadband wireless technologies (i.e., IEEE 802.11x and IEEE 802.16x). The next section provides a comparative analysis of the two wireless technologies. The paper then shows how the two technologies can be combined to provide broadband access. Lastly we conclude our paper.

## 2. Overview of Wi-Fi and WiMAX

### 2.1 IEEE 802.11 (Wi-Fi)

Wi-Fi stands for "wireless fidelity". However since most of our WLANs are based on those standards, the term Wi-Fi is used generally as a synonym for WLAN. Wi-Fi is a popular technology which allows any electronic device to exchange and transfer data wirelessly over the network giving rise to high speed internet connections. Any device which is Wi-Fi enabled (like personal computers, video game consoles, Smartphone, tablet etc.) can connect to a network resource like the internet through a wireless network access point. Now such access points also known as hotspots have a coverage area of about 20 meters indoors and even a greater area range outdoors, this is achieved by using multiple overlapping access points (Chan, 2005),(Intel Corp,2003).
However with all such features, Wi-Fi also suffers from certain shortcomings. Wi-Fi is known to be less secure than wired connections (such as Ethernet) because an intruder does not need a physical connection. Web pages that use SSL are secure but unencrypted internet access can easily be detected by intruders. Because of this, Wi-Fi has adopted various encryption technologies. The early encryption WEP, proved easy to break. Higher quality protocols (WPA, WPA2) were added later on. An optional feature added in 2007, called Wi-Fi Protected Setup (WPS) was deployed, but it also had a serious flaw that allowed an attacker to recover the router's password. The Wi-Fi Alliance has since updated its test plan and certification program to ensure all newly certified devices resist attacks. But security is still a major concern (Cam-Winget, et al., 2003), (Chandra Shekar, et al., 2005-2008).There are three well known 802.11 wireless family standard widely used today.

### 2.1.1 The IEEE 802.11b

A refined standard for the original 802.11 and was successful due to its high data rates of 11 Mb/s - range of 100 m to a maximum of a few hundred meters, operates on 2.4 GHz unlicensed band. 802.11b is the most widely deployed wireless network within the 802.11 wireless families (Cam-Winget, et al., 2003). It uses the DSSS modulation technique that is more reliable than the FHSS.

### 2.1.2 The IEEE 802.11g

The IEEE 802.11g wireless standard also operates on the 2.4 GHz band and has similar range and characteristics as the 802.11b. It has a data rate of 54Mbps (Xu, et al., 2006). The 802.11g has backward compatibility with 802.11b and differs only on the modulation technique; it uses Orthogonal Frequency Division Multiplexing (OFDM). This then makes the 802.11b devices not able to pick the signal from the 802.11g devices (Morrow, 2004).

### 2.1.3 The IEEE 802.11a

It operates in the 5GHz band with a maximum data rate of 54Mbps. The major disadvantage in deploying 802.11a with the other 802.11 standards b and g is that, they cannot co-exist, as they operate on different frequency bands (ProCurve Networking, 2005). 802.11b/g operates on the 2.4 GHz spectrum. There are some wireless card and access points which are compatible to all the three standards thereby supporting both the 2.4GHz and 5GHz frequencies band (Rensburg, 2006).

## 2.2 IEEE 802.16 (WiMAX)

WiMAX stands for “World Interoperability for Microwave Access”. It is a standard typically based on global interoperability including ETSI HIPERMAN, IEEE 802.16d-2004 for fixed, and 802.16e for mobile high-speed data. WiMAX is gaining popularity as a technology which delivers carrier-class, high speed wireless broadband at a much lower cost while covering large distance than Wi-Fi (Cam-Winget, et al., 2003). It has been designed to be a cost effective way to deliver broadband over a large area. It is intended to handle high-quality voice, data and video services while offering a high QoS (Westech Comms Inc., 2010).

WiMAX operates in between 10 and 66 GHz Line of Sight (LOS) at a range up to 50 km (30 miles) and 2 to 11GHz non Line-of-Sight (NLOS) typically up to 6 - 10 km (4 - 6 miles) for fixed customer premises equipment (CPE). Both the fixed and mobile standards include the licensed (2.5, 3.5, and 10.5 GHz) and unlicensed (2.4 and 5.8 GHz) frequency spectrum. However, the frequency range for the fixed standard covers 2 to 11 GHz while the mobile standard covers below 6 GHz. Depending on the frequency band, it can be Frequency Division Duplex (FDD) or Time Division Duplex (TDD) configuration. The data rates for the fixed standard will support up to 75 Mbps per subscriber in 20 MHz of spectrum, but typical data rates will be 20 to 30 Mbps. The mobile applications will support 30 Mbps per subscriber, in 10 MHz of spectrum, but typical data rates will be 3 - 5 Mbps.

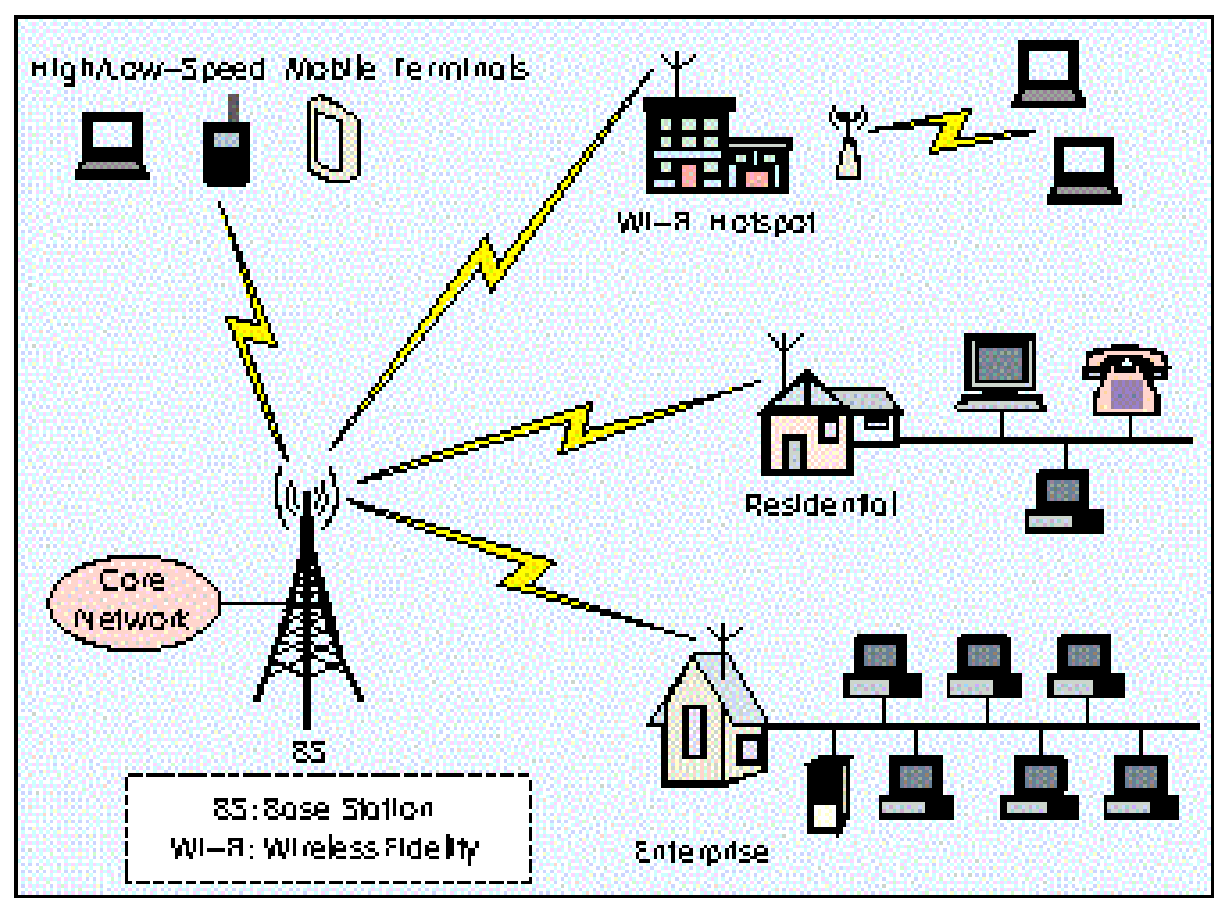

**Figure 1**

Applications of fixed WiMAX (802.16-2004) include wireless E1 enterprise backhaul and residential ‘last mile’ broadband access, while applications for mobile WiMAX (802.16e) include nomadic and mobile consumer wireless DSL service. Other WiMAX applications include: connecting Wi-Fi hotspots with each other and to other parts of the Internet; providing a wireless alternative to cable and DSL for last mile (last km) broadband access. On flexibility, WiMAX can be deployed in any terrain across all geographical areas.

## 2.2 PHY (Physical) Layer

Apart from the usual functions such as randomization, forward error correction (FEC), interleaving, and mapping to QPSK and QAM symbols, the standard also specifies optional multiple antenna techniques. This includes space time coding (STC); beam forming using adaptive antennas schemes, and multiple input multiple output (MIMO) techniques which achieve higher data rates. The OFDM modulation/demodulation is usually implemented by performing fast Fourier transform (FFT) and inverse FFT on the data signal. The MAC layer used by WiMAX is based on a time division multiple access (TDMA) mechanism to allow a homogeneous distribution of the bandwidth between all the devices which is more effective and support several channels compared to the mechanism used by Wi-Fi (CSMA-CA). This makes it possible to obtain a better optimization of the radio spectrum with better efficiency (bits/seconds/Hertz). Thus, WiMAX has an efficiency of 5 Bps/Hz compared to the 2.7Bps/Hz of Wi-Fi that makes it possible to transmit 100 Mb/s on 20 MHz channel.

## 3. Wi-Fi and WiMAX: Comparison

WiMAX is different from Wi-Fi in many respects. In fact, Wi-Fi can operate at distances as great as WiMAX but there are two reasons why it doesn't. One of the reasons is that radios operating in the unlicensed frequencies are not allowed to be as powerful as those operated with licenses; and from convention, less power means less distance. These regulations are based on the dated assumption that devices can't regulate themselves — but the assumption may be correct over great enough distances. The second reason as to why Wi-Fi access points don't serve as wide an area as WiMAX access points do is the common engineering belief that the problem of everybody shouting at once, even if it's surmountable in a classroom, would be catastrophic in a larger arena. The Wi-Fi MAC layer uses contention access. This causes users to compete for data throughput to the access point. Wi-Fi even has problems with interference, and throughput and that is why triple play (voice, data, and video) technologies cannot be hosted on traditional Wi-Fi. In contrast, 802.16 use a scheduling algorithm (Xu, et al., 2006). This algorithm allows the user to only compete once for the access point. This gives WiMAX inherent advantages in throughput, latency, spectral efficiency, and advanced antenna support. From the technical point of view, it can be seen that both of these two wireless technologies are not basically addressed at the same market but are very complementary. Wi-Fi is basically an implementation of wireless local area network within a short range like a small building, a college or an institutional campus. WiMAX on the other hand is a metropolitan technology whose objective is to interconnect houses, buildings or even hot spots to allow communication between them and with other networks.

Although not being targeted on the same use, more recently WiMAX technology has several advantages compared to Wi-Fi. Such as: a better reflection tolerance; a better penetration of obstacles; and an increased in the number of interconnections (a few hundreds of equipment rather than some tens of equipment for Wi-Fi). It's obvious that the WiMAX standard goal is not to replace Wi-Fi in its applications but rather to supplement it in order to form a wireless network web. Despite the similarity in equipment cost, WiMAX technology requires a costly infrastructure in contrast to Wi-Fi which can easily be installed using low cost access points. These two wireless technologies have common components in their operations with a major difference in the communication range. The following table 1 gives the detailed comparative analysis of the two broadband wireless access networks (Wi-Fi and WiMAX).

## 4. Wi-Fi and WiMAX: Application

Both Wi-Fi and WiMAX can be integrated and overlay. If they can be integrated, it means that WiMAX and Wi-Fi will support each other. Both of them will be synergized to serve bigger and many more subscribers. WiMAX and Wi-Fi can offer some potentially significant cost savings for mobile network operators by providing an alternate means to backhaul BS traffic from cell site to the BS controllers. Mobile network operators typically utilize some type of wired infrastructure that they must buy from an incumbent operator. A Wi-Fi or WiMAX mesh can offer a much more cost-effective backhaul capability for BSs in metropolitan environments. Using Wi-Fi and WiMAX open broadband wireless standards and implementing mobile computing, governments and partners can quickly and cost-effectively deploy broadband to areas not currently served, with little or no disruption to existing infrastructures (Chandra Shekar, et al., 2005-2008).

Standards-compliant WLANs and proprietary Wi-Fi mesh infrastructures are being installed rapidly and widely throughout the world. Standards-compliant WiMAX products can provide NLOS backhaul solutions for these local networks and WiMAX subscriber stations can currently provide Internet access to customers such as schools and other educational institutions and campuses. If they can be made to overlap in coverage they can be functioned to support each other (if they were in one operator) and will be opponent if they were in different operators. Various configurations that can be applied by WiMAX and Wi-Fi operators if they were integrated are as follows:

**Table 1** Comparison between Wi-fi & WiMAX

| **Feature** | **WiMAX (802.16a)** | **Wi-Fi (802.11b)** | **Wi-Fi (802.11a/g)** |
|---|---|---|---|
| **Primary Application** | Broadband Wireless Access | Wireless LAN | Wireless LAN |
| **Frequency Band** | Licensed/Unlicensed 2 G to 11 GHz | 2.4 GHz ISM | 2.4 GHz ISM (g) 5 GHz U-NII (a) |
| **Channel Bandwidth** | Adjustable 1.25 M to 20 MHz | 25 MHz | 20 MHz |
| **Half/Full Duplex** | Full | Half | Half |
| **Radio Technology** | OFDM (256-channels) | Direct Sequence Spread Spectrum | OFDM (64-channels) |
| **Bandwidth Efficiency** | <=5 bps/Hz | <=0.44 bps/Hz | <=2.7 bps/Hz |
| **Modulation** | BPSK, QPSK, 16-, 64-, 256-QAM | QPSK | BPSK, QPSK, 16-, 64-QAM |
| **FEC** | Convolutional Code Reed-Solomon | None | Convolutional Code |
| **Encryption** | Mandatory- 3DES Optional-AES | Optional- RC4 (AES in 802.11i) | Optional- RC4 (AES in 802.11i) |
| **Mobility** | Mobile WiMAX (802.16e) | In development | In development |
| **Mesh** | Yes | Vendor Proprietary | Vendor Proprietary |
| **Access Protocol** | Request/Grant | CSMA/CA | CSMA/CA |

## A. BACKHAUL

The configuration is shown in fig. 2.By combining the two technologies, WiMAX functioning as a backhaul while Wi-Fi connected directly to the subscriber.

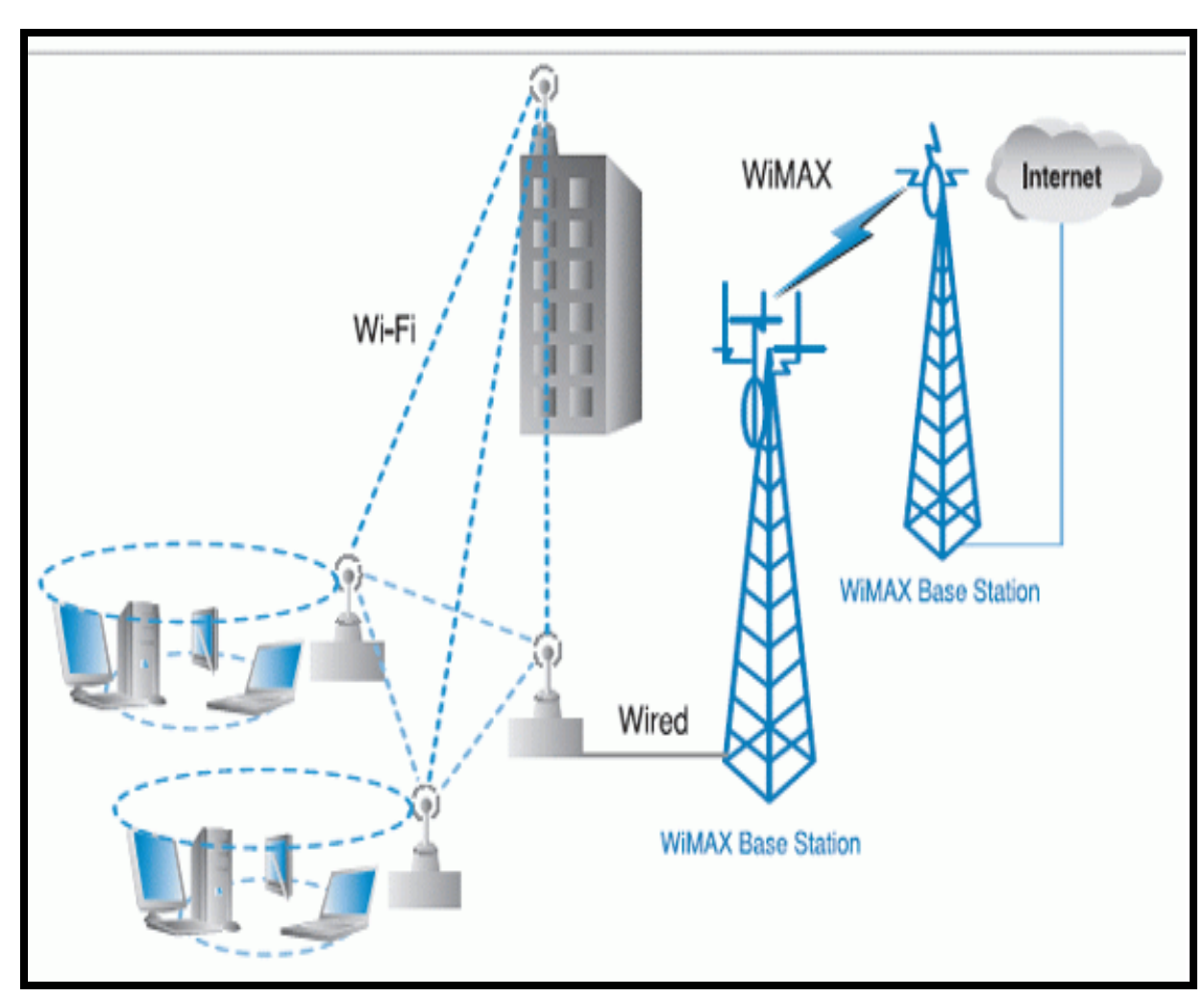

**Figure 2** Backhaul Architecture

## B. BACKHAUL INTER WI-FI MESH NETWORK

The configuration is shown in fig. 3. In this step, WiMAX has been used directly as a part of Wi-Fi Mesh Network. Subscriber Terminal of WiMAX is put on access point of Wi-Fi Mesh Network so that the Wi-Fi network automatically will be more reliable in wider coverage area and reduce cost connection that is caused by cable drawing in each AP installation. The solution principally can increase performance and robustness of the Wi-Fi network.

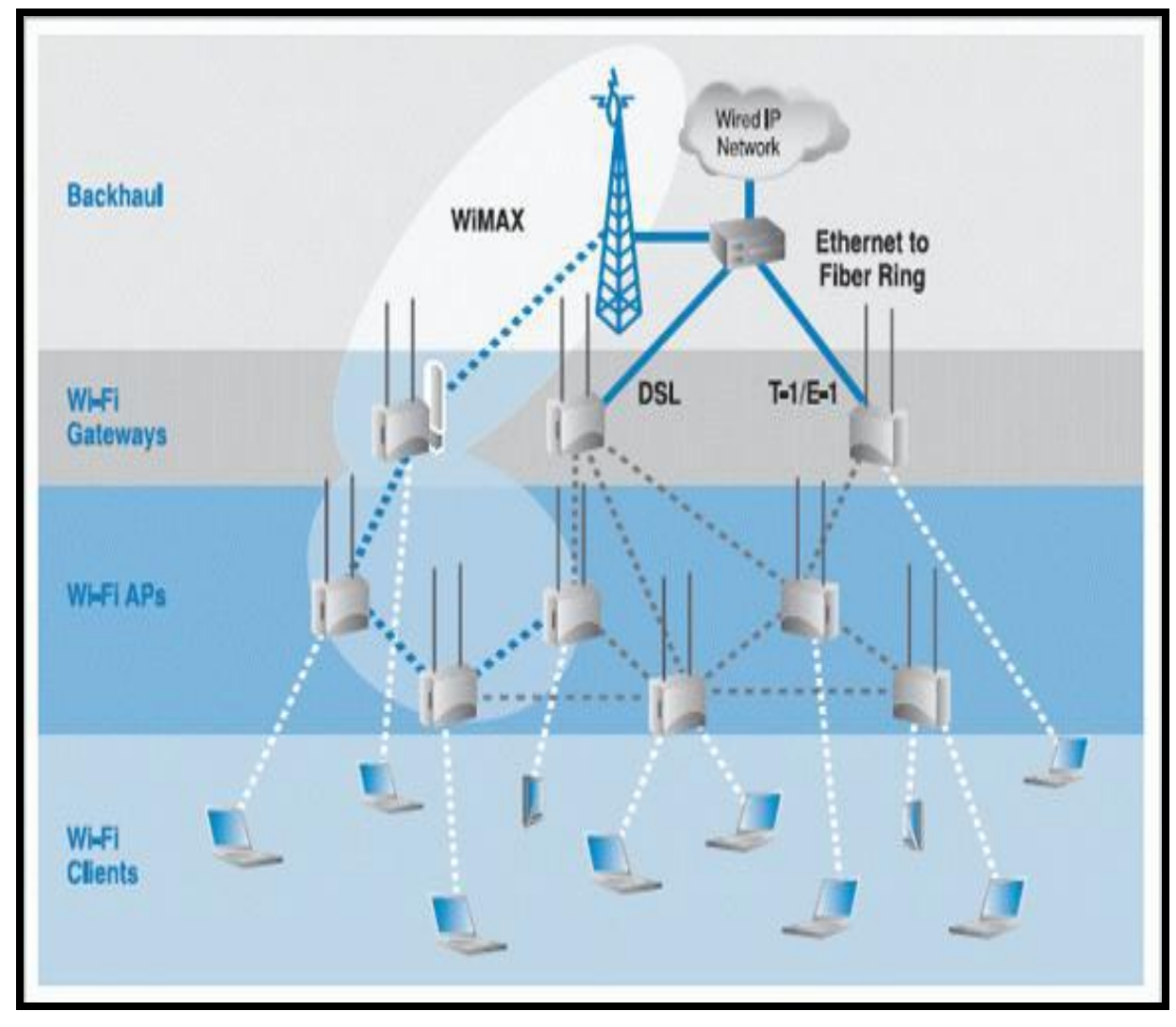

**Figure 3** Backhaul Inter Wi-Fi Mesh Network

### C. Wi-Fi-WiMAX FULL INTEGRATED

Figure 4 shows another combination between Wi-Fi and WiMAX. Here in this case, communication can be done up to client level. WiMAX coverage is overlapping with Wi-Fi coverage. It gives better service choices, more flexible to the changes of network and is more user friendly with connection ease compatible with terminal that has been owned. Moreover with dual AP radio implementation (Wi-Fi and WiMAX), integration will be easier and network development also can be faster.

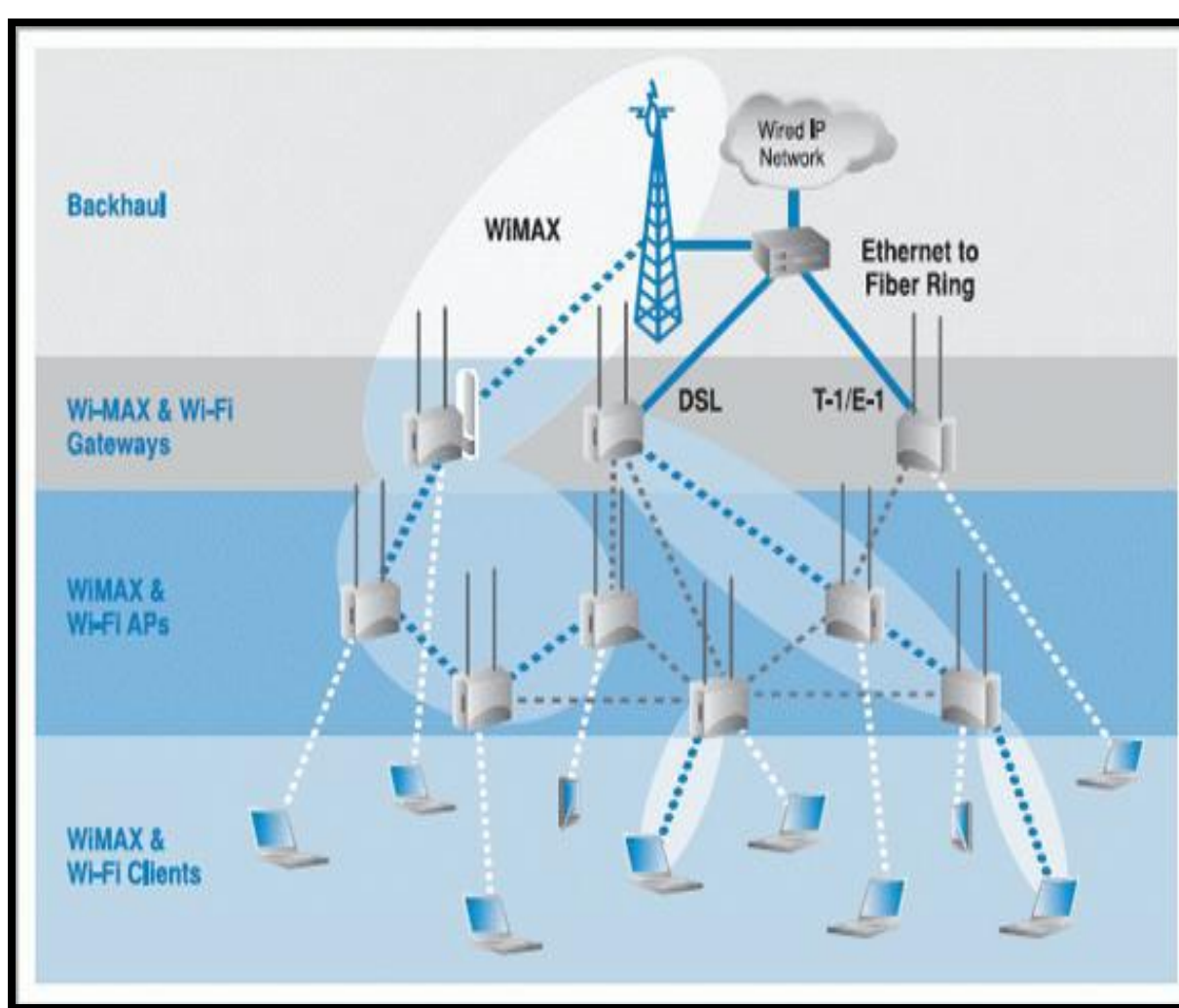

**Figure 4** Wi-Fi – WiMAX Fully Integrated Architecture

Combination of both these technologies in any of the three configurations shown and discussed above gives sufficient solution, especially for data communication system that is a still problem nowadays.

## 5. CONCLUSION

This paper has presented a precise description of two of the most prominent developing wireless access networks and even discussed as to how these technologies may collaborate together to form an alternatives for implementing last-mile wireless broadband services. Detailed technical comparative analysis between the 802.11 (Wi-Fi) and 802.16 (WiMAX) wireless networks that provide alternative solution to the problem of information access in remote inaccessible areas where wired networks are not cost effective has been looked into. This work has proved that the WiMAX standard goal is not to replace Wi-Fi in its applications but rather to supplement it in order to form a wireless network web.

## REFERENCES

1. Rensburg, J.J, "Investigation of the Deployment of 802.11 Wireless Networks", M.Sc Thesis. University of Rhodes: Ghramstown, South Africa (2006).
2. Chan, H. Anthony. "Overview of Wireless Data Network Standards and Their Implementation Issues." Talk presented at the 12th ICT Cape Town (2005).
3. Intel Corp, "IEEE 802.16 and WiMAX: Broadband Wireless Access for Everyone", [Online] Available http:// www.intel.com/ebusiness/pdf/wireless/intel/80216_wimax.pdf (2003).
4. Black Box, "802.11: Wireless Networking", White Paper, [Online] Available http://www.blackbox.com/Tech_Support/White Papers/802.11-Wireless-Networking2.pdf (2005).
5. Cam-Winget, Nancy, et al. "Security flaws in 802.11 data link protocols."Communications of the ACM 46.5 (2003) pp. 35-39.
6. Morrow, R.,"Wireless Network Coexistence", McGraw-Hill: New York, NY (2004).
7. ProCurve Networking, "Planning a Wireless Network", White Paper, Hewlett-Packard Development Company: Los Angeles, CA. [Online] Available www.hp.com/rnd/pdfs/802.11 11technicalbrief.pdf (2004).
8. Xu, Sen, Manton Matthews, and Chin-Tser Huang. "Security issues in privacy and key management protocols of IEEE 802.16." In Proceedings of the 44th annual South east regional conference, ACM, (2006) pp. 113-118.
9. Westech Comms Inc., "Can WiMAX Address Your Application", White Paper (2006).
10. D.V. Chandra Shekar, V. J., "Wireless security: A comparative analysis for the next generation networks", Journal of Theoretical and Applied Information Technology (2005-2008), pp. 822-831.